\documentstyle[aps,multicol,psfig,eqsecnum]{revtex}  
\input epsf
\def\deltat{r_0}               

\def\inte{g}                   

\def\dmhrs{{\cal D} \Omega }

\def \ofield#1{ \Omega(\hat #1)}
\def\HRS{{\rm HRS}}
\def\LRS{{\rm LRS}}
\def\WRS{{\rm 1RS}}
\def\ZRS{{\rm 0RS}}
\def\action{{\cal S}}
\def\perdef{{\cal P}}

\newcommand{\oofield} [2] {\Omega({\hat #1}_{#2})}

\def\be{\begin{equation}}
\def\ee{\end{equation}}
\def\ba{\begin{eqnarray}}
\def\ea{\end{eqnarray}}

\def\coeLVzero{A}
\def\coeLVtwo{B}
\def\coeLVthree{C}
\def\coeffzero{\coeLVzero_{n}}
\def\coefftwo{\coeLVtwo_{n}}
\def\coeffthree{\coeLVthree_{n}}
\def\bml{\begin{mathletters}}
\def\eml{\end{mathletters}}
\newcommand{\eqbreak}{
\end{multicols}
\widetext
\noindent
\rule{.48\linewidth}{.1mm}\rule{.1mm}{.1cm}
}
\newcommand{\eqresume}{
\noindent
\rule{.52\linewidth}{.0mm}\rule[-.1cm]{.1mm}{.1cm}\rule{.48\linewidth}{.1mm}
\begin{multicols}{2}
\narrowtext
}
\begin{document} 
\draft 
\title{Connecting the vulcanization transition to percolation}
\author{Weiqun Peng$^1$, Paul M.~Goldbart$^1$ and Alan J.~McKane$^{1,2}$} 
\address{
$^1$Department of Physics, 
University of Illinois at Urbana-Champaign,
1110 West Green Street, 
Urbana, Illinois 61801, U.S.A.}
\address{
$^2$Department of Theoretical Physics, 
University of Manchester,
Manchester M13 9PL, UK}
\maketitle
\begin{abstract} 
The vulcanization transition is addressed via a minimal 
replica-field-theoretic model.  The appropriate long-wave-length 
behavior of the two- and three-point vertex functions is considered 
diagrammatically, to all orders in perturbation theory, and identified 
with the corresponding quantities in the Houghton-Reeve-Wallace 
field-theoretic approach to the percolation critical phenomenon.  Hence, 
it is shown that percolation theory correctly captures the critical 
phenomenology of the vulcanization transition associated with the 
liquid and critical states. 
\end{abstract}
%
%
%
\noindent
\section{Introduction and overview}
\label{SEC:Intro}
The vulcanization transition (henceforth denoted VT) is the generic 
equilibrium phase transition from a parent liquid state of matter to 
an amorphous solid state, driven by the imposition of a sufficient 
density of permanent random constraints between the constituents of 
the liquid.  In the most common setting of the VT, the constituents 
of the liquid are macromolecules, the locations of which provide the 
annealed random (i.e.~thermally equilibrating) variables.  The 
constraints are commonly provided by covalent chemical bonds 
(i.e.~cross-links), and impart quenched randomness upon the system. 
Over the past few years, a rather detailed description of the VT has 
been developed, ranging from a mean-field theory of the emergent 
amorphous solid state (including its structural~\cite{epl,univ} 
and elastic properties~\cite{REF:elasticity}, and its 
stability~\cite{REF:stability}; 
for reviews, see Refs.~\cite{cross,REF:Trieste}) 
to the critical properties of the VT itself~\cite{REF:WPandPG}.  

The present Paper aims to extend the description of the 
critical properties of the VT by exploring its relationship 
with the percolation transition (henceforth denoted PT; 
for a review, see Ref.~\cite{REF:PercRev}). 
This relationship, which has long been anticipated on physical 
grounds~\cite{REF:PercAnticip,REF:PercWork}, has recently found 
support both at the mean-field level~\cite{epl,cross}, and beyond, via a 
renormalization-group approach~\cite{REF:WPandPG}.  Specifically, it was 
recently shown that the order-parameter correlator near the VT (which 
probes for relative localization of particles) and its physical analog 
in PT (viz., the connectedness function) are governed by the same 
critical exponents, at least to first order in an expansion about 
the upper critical dimension six~\cite{REF:WPandPG}.

The central result of the present Paper is the explicit reduction of 
certain basic critical properties of the VT to equivalent basic 
critical properties of the PT~\cite{REF:APS2001}.  As we shall see, 
this reduction can be accomplished via an exact diagrammatic analysis 
of the complete perturbative expansion of the appropriate vertex 
functions of the VT.  These are shown to furnish, in the replica limit, 
precisely the field-theoretic formulation of the PT due to 
Houghton, Reeve and Wallace~\cite{REF:HRW} (which henceforth we shall 
refer to as HRW).  Hence, we establish that the critical properties of 
the VT and the PT are identical, not just to first order but to all 
orders in the departure of the spatial dimension $d$ from the upper 
critical dimension.

It is worth observing that the VT and the PT do, nevertheless, represent 
distinct physical phenomena.  This is exemplified, e.g., by the amorphous 
solid state that emerges at the VT, which does not have an evident 
counterpart in PT.  Another point of distinction is 
revealed by the role of fluctuations in low-dimensional systems, 
which are expected to have qualitatively different impacts on the states 
emerging at the VT and the PT~\cite{REF:LowDim}.  Yet another point 
of distinction concerns the nature of the degrees of freedom involved 
in the description of the VT and the PT.  The former arises in systems 
having both quenched and equilibrating randomness, whereas the latter 
takes place in systems involving just one type of randomness (typically 
taken to be the quenched randomness); see Ref.~\cite{REF:LowDim}.  

After completion of the present work we learned of the elegant work 
of Janssen and Stenull~\cite{REF:JandS}, conducted independently of and 
simultaneously with the present work, which builds on earlier work on 
random resistor networks and percolation to arrive at, inter alia, 
essentially the same results as those contained in the present paper 
via a related approach.
\section{Minimal model of the vulcanization transition}
\label{SEC:Models}
Our analysis is based upon a minimal model of the VT that accounts 
for thermal fluctuations in the positions of the constituents of the 
parent liquid, short-range repulsions between these constituents, and 
permanent random constraints (e.g.~resulting from cross-linking) which 
explicitly reduce the collection of configurations accessible to the 
constituents.  This minimal model yields a rich mean-field picture of 
the structure and elastic response of the amorphous solid state, the 
former aspect having been verified by the computer simulations of Barsky 
and Plischke~\cite{REF:SJB_MP}. 

The minimal model for the VT can be built (in the spirit of the 
Landau-Wilson scheme for continuous phase transitions) on the general
basis of symmetry considerations and a gradient expansion~\cite{univ}.  
The appropriate order parameter $\Omega$, whose expectation value 
detects the emergence of the amorphous solid state, has been discussed 
elsewhere~\cite{REF:pointer}.  The quenched random constraints are 
accounted for via the replica technique, which incorporates the Deam-Edwards 
model~\cite{REF:DeamEd} for their statistics (viz.~the statistics of 
the random constraints are determined by the instantaneous correlations of 
the unconstrained system).  The additional replica associated with the 
Deam-Edwards model leads to a situation in which one considers the $n\to 0$ 
limit of a system of not of $n$ but of $n+1$ replicas.  

\begin{figure}[hbt]
\epsfxsize=2.5in
\centerline{\epsfbox{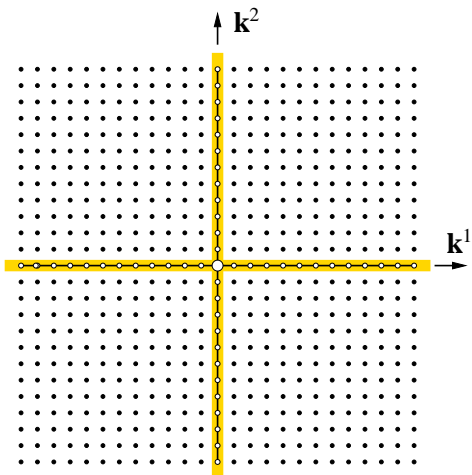}}  
\vskip0.50cm
\caption{Decomposition of the space of replicated wave vectors. 
Off-axis wave vectors lie in the \HRS; 
on-axis (but off-origin) wave vectors lie in the \WRS; 
the wave vector at the origin is the \ZRS.
\label{FIG:Rs_decomp}}
\end{figure}

The resulting Landau-Wilson effective Hamiltonian takes the form of a 
cubic field theory involving the field $\Omega$, the argument of this 
field lying in $(n+1)$-fold replicated $d$-dimensional space:
\begin{equation}
\action
\big(\Omega\big)=
\frac{1}{V^{n+1}}\frac{1}{2}\sum_{\hat{k}\in{\HRS}}
\Big(\deltat+\hat{k}\cdot\hat{k}\Big)
\big\vert\ofield{k}\big\vert^{2}+
\,\,\,
\frac {1}{V^{2(n+1)}} \frac {\inte}{3!}
        \!\!\!\!
\sum_{\,\,\,\,\hat{k}_1,\hat{k}_2,\hat{k}_3\in\HRS}
        \!\!\!\!
\oofield{k}{1}\,
\oofield{k}{2}\,
\oofield{k}{3}\,
\delta_{{\hat{k}_1}+{\hat{k}_2}+{\hat{k}_3},{\hat{0}}}\,.
\label{EQ:LG_longwave}
\end{equation}
The free-energy density $f$ is (up to uninteresting factors that 
we shall ignore) related to this Hamiltonian via  
$f\propto\lim_{n\to 0}n^{-1}{\ln\{\int\dmhrs \exp(-\action)\}}$; 
the functional integral is taken over 
the independent components of $\Omega$ that 
feature in $\action$.  In $\action$, the quantity $\deltat$ 
is the VT {\it control parameter\/} which, near the VT, is linearly 
related to the density of random constraints.  (To ease comparison 
with the HRW field theory of the PT, the coefficients and fields 
featuring in $\action$ are not defined exactly as they have been 
in our earlier works~\cite{REF:Notation}.)\thinspace\ 
The symbol ${\hat k}$ denotes the replicated wave vector 
$\{{\bf k}^0,{\bf k}^1,\ldots,{\bf k}^n\}$; the extended 
scalar product $\hat{k}\cdot\hat{c}$ is denoted 
${\bf k}^0\cdot{\bf c}^0+
 {\bf k}^1\cdot{\bf c}^1+
 \cdots+{\bf k}^n\cdot{\bf c}^n$. 
The specification \HRS\ arises from the following considerations.
Consider the space of replicated wave vectors $\hat{k}$.
We decompose this space into three disjoint sets: 
(i)~{\it the higher replica sector\/} \HRS, 
which consists of those $\hat{k}$ containing at least two 
nonzero component-vectors ${\bf k}^\alpha$;  
(ii)~{\it the one replica sector\/} \WRS, 
which consists of those $\hat{k}$ containing 
exactly one nonzero component-vector ${\bf k}^\alpha$; and 
(iii)~{\it the zero replica sector\/} \ZRS, 
which consists of the vector ${\hat k} = \hat 0$. 
This decomposition is illustrated schematically in 
Fig.~\ref{FIG:Rs_decomp} for the case of two replicas. 
It is especially straightforward to visualize this 
decomposition if the volume of the system is kept finite 
(and periodic boundary conditions are imposed) so that replicated 
plane waves with discrete, equally spaced, replicated wave vectors 
provide the natural complete set of functions.
The symbol $\sum_{\hat{k}\in{\HRS}}$ indicates a summation over the 
replicated wave vector $\hat{k}$, subject to the restriction that 
$\hat{k}$ lies in \HRS.  This condition on the summation over $\hat{k}$ 
is essential.  Physically, it reflects the fact that no macroscopically 
inhomogeneous modes (such as crystalline modes) order or fluctuate 
critically in the vicinity of the VT, such modes being stabilized by the 
excluded-volume interactions.  Mathematically, this condition reduces 
the symmetry of the theory from one that contains the rotation group in 
replicated $(n+1)d$-dimensional space to one that contains only 
rotations {\it within\/} each individual replica, along with 
the permutation of the replicas.
\section{Demonstrating the equivalence of the critical properties 
of the vulcanization and percolation transitions}
\label{SEC:Relationship}
\subsection{Overall strategy}
\label{SEC:strategy}
We now explain the strategy that we shall use to relate the 
VT and the PT.  We shall focus on the replica limit of the 
long-wave-length behavior of the two- and three-point vertex 
functions, 
$\Gamma_n^{(2)}(\hat k)$ and 
$\Gamma_n^{(3)}({\hat k}_1,{\hat k}_2)$ 
in the VT field theory. 
The physical significance $\Gamma_n^{(2)}(\hat k)$ as a probe of 
connectedness has been elucidated in Ref.~\cite{REF:WPandPG}. Now, the 
symmetry of the VT field theory dictates that the only suitably invariant 
term quadratic in the wave vector $\hat{k}$ is $\hat{k}\cdot\hat{k}$.
Thus, in a long-wave-length expansion for $\Gamma_n^{(2)}(\hat k)$, 
we have 
\bml
\begin{eqnarray}
\Gamma_n^{(2)}({\hat k})
&=&
\Gamma_n^{(2)}({\hat 0})+
{1\over{(n+1)d}}
\sum_{\alpha=0}^{n}
\left(
{\partial\over{\partial{\bf k}^{\alpha}}}\cdot
{\partial\over{\partial{\bf k}^{\alpha}}}
\right)
\Gamma_n^{(2)}(\hat k)\Big\vert_{{\hat k}={\hat 0}}
\,\,{\hat k}\cdot{\hat k}+\cdots
\\
&=&
\coeffzero+
\coefftwo
{\hat k}\cdot{\hat k}+\cdots\,\,.
\label{EQ:VFTexp}
\end{eqnarray}%
\eml%
As the upper critical dimension for the VT is six, and this 
is the dimension about which one may imagine expanding, 
general renormalizability considerations demand that just 
these two vertex functions 
($\Gamma_n^{(2)}$ and $\Gamma_n^{(3)}$)
contain the primitive divergences, 
and do so via the constants $\coeffzero$, $\coefftwo$ and 
$\coeffthree\equiv\Gamma_n^{(3)}({\hat 0},{\hat 0})$ 
(see Ref.~\cite{REF:Wavevectorzero}).

Having identified the quantities central to a renormalization-group 
analysis of the VT, we shall establish that these quantities are 
identical, in the replica limit, to the corresponding quantities 
in percolation theory.  To do this we shall make use of a convenient 
representation of the critical properties of percolation theory, viz., 
the HRW field theory representation~\cite{REF:HRW}.  So that we know 
what we need to make contact with, we pause to give a brief account 
of this HRW representation, the Landau-Wilson Hamiltonian for which 
is given by 
\begin{equation}
{\cal H}=\int d^d x 
\left\{ \frac{1}{2} (\nabla\phi)^2 - 
\frac{1}{2} (\nabla\psi)^2 + 
\frac{1}{2} r_0(\phi^2- \psi^2) +
\frac{g}{3!}(\phi+\psi)^3\right\}, 
\label{EQ:HRW_Action}
\end{equation}
where $\phi$ is an ordinary field but $\psi$ is a ghost field.  As HRW 
have shown, provided one enforces the rule that {\it only graphs that 
are connected by $\phi$-lines are included\/}, the two- and three-point 
$\phi$ vertex functions are identical (order by order in perturbation 
theory in the coupling constant $g$) to those of the one-state 
(i.e.~percolation) limit of the Potts model.  We mention, in passing,
that this HRW representation consists of fields residing on 
$d$-dimensional space, and does not necessitate the taking of a 
replica (or Potts) limit.  However, it does require the additional 
rule by which certain diagrams are excluded by hand. 

Our strategy is as follows.  
Consider the standard Feynman diagram expansion for the two- and 
three-point vertex functions of the VT field theory in powers of the 
coupling constant $\inte$ in the Hamiltonian~(\ref{EQ:LG_longwave}). 
\hfil\break\noindent
(i)~To deal with the constraint that the internal wave vectors in the 
resulting diagrams reside in the \HRS, we relax this constraint 
on summations over internal wave vectors but compensate for this by 
making appropriate subtractions of terms. 
\hfil\break\noindent
(ii)~Next, we observe that all diagrams for the two- and three-point 
vertex functions can be organized into two categories: those in 
which there is at least one route between every pair of external 
points via propagators having unconstrained wave vectors (which 
we call {\it freely-connected diagrams\/}); and the remaining diagrams, 
in which there is at least one pair of external points between which 
no paths exist consisting solely of propagators having unconstrained 
wave vectors (which we call {\it freely-unconnected diagrams\/}).  
Having made this categorization, we show that the appropriate version 
of wave-vector conservation renders the freely-unconnected diagrams 
negligible in the thermodynamic limit, leaving us with a representation 
that is already reminiscent on the HRW approach.
\hfil\break\noindent
(iii)~At this stage we have reduced the construction of the two- and 
three-point vertex functions to the computation of freely-connected 
diagrams only.  Next, via a straightforward combinatorial analysis, 
we show that, in the replica limit, only a small class of diagrams 
survive. 
\hfil\break\noindent
(iv)~Finally, we explain how, again in the replica limit, the values of the 
remaining diagrams are precisely those occurring in the HRW prescription 
for percolation.
\hfil\break\noindent
We now set about implementing this strategy.
\subsection{Relaxing the constraint to 
higher replica sector wave vectors}
\label{SEC:relax}
In the VT field theory the internal wave vectors occurring 
in the Feynman diagrams are 
constrained to lie in the \HRS.  In order to perform summations 
over these wave vectors, it is convenient to  work with the continuum 
of wave vectors (i.e.~to take the thermodynamic limit) rather than 
the discrete lattice of them.  In order to be able to take this 
limit, we re-express summations over \HRS\ wave vectors  
in terms of unconstrained summations over  $(n+1)d$-dimensional  
wave vectors, together with further unconstrained summations over 
$d$-dimensional wave vectors (and certain trivial additional terms).
To do this, we note that for a generic function $F(\hat k)$ we have
\bml
\begin{eqnarray}
\sum_{\hat{k}\in{\HRS}}F(\hat k)
&=&
\sum_{\hat{k}}
\left(
1-
\Big(
\sum_{\alpha=0}^{n}
\sum_{{\bf q}\neq{\bf 0}}  
\delta_{{\bf k}^{\alpha},{\bf q}}
\prod_{\beta(\ne\alpha)}
\delta_{{\bf k}^{\beta},{\bf 0}}
\Big)-
\delta_{\hat{k},\hat {0}}
\right)F(\hat k)
\\
&=&
\sum_{\hat{k}}
\left(
1-
\Big(
\sum_{\alpha=0}^{n}
\sum_{{\bf q}}  
\delta_{{\bf k}^{\alpha},{\bf q}}
\prod_{\beta(\ne\alpha)}
\delta_{{\bf k}^{\beta},{\bf 0}}
\Big)-
n\delta_{\hat{k},\hat {0}}
\right)F(\hat k)
\\
&=&
\sum_{\hat{k}}F(\hat k)-
\sum_{\alpha=0}^{n}
\sum_{{\bf q}^{\alpha}}
F({\bf 0},\ldots,{\bf 0},{\bf q}^{\alpha},{\bf 0},\ldots)
-nF(\hat 0), 
\label{EQ:Corr_Decomposition}
\end{eqnarray}%
\eml%
which effects the re-expression of the summations just described.  Note 
that, as it always comes with the factor $n$, the 
$\delta_{\hat{k},\hat {0}}$ term will vanish in the replica limit, 
and can therefore be safely 
ignored.  We shall refer to the wave vectors included in the term
$\sum_{\alpha}\sum_{{\bf q}}$ as {\it lower replica sector\/} (\LRS) 
wave vectors.  Via these steps one can relax a constrained summation 
over \HRS\ wave vectors, instead freely summing over all replicated
wave vectors, provided one compensates by augmenting the summand with the 
factor 
\begin{equation}
1-\sum_{\alpha=0}^{n}\sum_{{\bf q}}  
\delta_{{\bf k}^{\alpha},{\bf q}}\prod_{\beta(\ne\alpha)}
\delta_{{\bf k}^{\beta},{\bf 0}}\,\,.
\label{EQ:conrelax}
\end{equation}

\begin{figure}[hbt]
\epsfxsize=1.4in
\centerline{\epsfbox{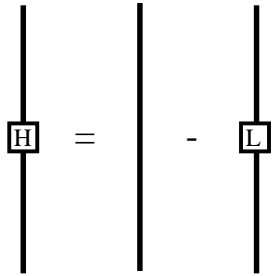}}  
\vskip0.50cm
\caption{Decomposition of an \HRS\ propagator (indicated by H)
into an unconstrained propagator (the unadorned line), 
less an \LRS\ propagator (indicated by L).
\label{FIG:Correlator_Decomposition}}
\end{figure}

How does this constraint relaxation manifest itself in the  
setting of Feynman diagram computations?  One simply augments 
every internal propagator $V^{n+1}G_0(\hat{k})$ with a 
factor~(\ref{EQ:conrelax}):
\bml
\begin{eqnarray}
G_0(\hat{k})\equiv
\frac{1}{\deltat+\hat{k}\cdot\hat{k}}
&\longrightarrow&
\frac{1-\sum_{\alpha}\sum_{{\bf q}}  
\delta_{{\bf k}^{\alpha},{\bf q}}\prod_{\beta(\ne\alpha)}
\delta_{{\bf k}^{\beta},{\bf 0}}}{\deltat+\hat{k}\cdot\hat{k}}
\\
&=&
\frac{1}{\deltat+\hat{k}\cdot\hat{k}}-
\frac{1}{\deltat+\hat{k}\cdot\hat{k}}
\sum_{\alpha=0}^{n}\sum_{{\bf q}}  
\delta_{{\bf k}^{\alpha},{\bf q}}
\prod_{\beta(\ne\alpha)}
\delta_{{\bf k}^{\beta},{\bf 0}}.
\label{EQ:Bare_Correlator}
\end{eqnarray}%
\eml%
How this decomposition is expressed diagrammatically is shown in 
Fig.~\ref{FIG:Correlator_Decomposition}.

In this manner, each \HRS\ internal line in a Feynman diagram can be 
decomposed into an unconstrained internal line, less an \LRS\ internal 
line.  Thus, the various vertex functions can be expressed in terms 
of Feynman diagrams composed of unconstrained lines together with as \LRS\ 
internal lines.  Note, for future reference, that physically meaningful vertex 
functions have external wave vectors in the \HRS.  

\begin{figure}[hbt]
\epsfxsize=3.5in
\centerline{\epsfbox{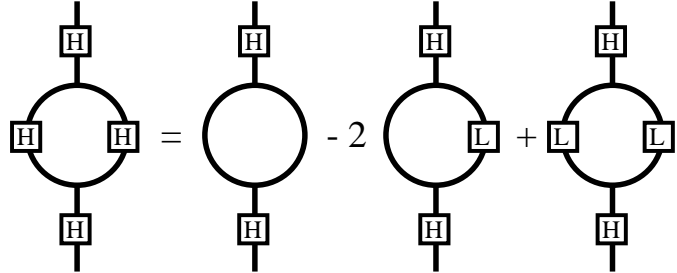}}  
\vskip0.50cm
\caption{Decomposition of the one-loop diagram for the two-point 
vertex function.  On the right-hand side of this equation, the 
first and second diagrams are freely-connected diagrams, and the 
third is a freely-unconnected diagram.
\label{FIG:Ex_Decomposition}}
\end{figure}

We illustrate this decomposition for the case of a simple diagram in 
Fig.~\ref{FIG:Ex_Decomposition}.
More generally, we arrive at the following modified Feynman rules 
for the VT field theory: 
\hfil\break\noindent
(i)~Write down all diagrams arising from the original theory. In these
diagrams all wave vectors are constrained to the \HRS.
\hfil\break\noindent
(ii)~Replace each diagram with the collection of diagrams obtained by 
allowing each internal line to carry either an unconstrained replicated 
wave vector or a \LRS\ wave vector.  (Thus, a diagram with $L$ internal 
lines spawns a total of $2^L$ diagrams.)\thinspace\  
Identically-valued diagrams can be represented by a single diagram, 
together with a suitable combinatorial factor (see, e.g., 
the factor of 2 in Fig~\ref{FIG:Ex_Decomposition}).
\hfil\break\noindent
(iii)~Provide a factor of $-1$ for each \LRS\ internal line. 
\hfil\break\noindent
At this stage we observe that the combinatorics of our diagrammatic 
expansion coincides with those of the HRW expansion, provided one 
identifies the internal unconstrained and \LRS\ lines of the VT 
theory with, respectively, the corresponding internal $\phi$- and 
$\psi$-lines of the HRW representation.

We have, however, yet to show that the diagrams removed by hand in 
HRW can be safely omitted from the VT theory, and that the numerical 
values of the (replica limits of) the VT diagrams are identical to 
those of the HRW diagrams.  We shall establish these facts in 
the following subsections.
\subsection{Elimination of freely-unconnected diagrams}
\label{SEC:trails}
We remind the reader that in the HRW theory for the two- and three-point 
$\phi$-field (i.e.~the physical) vertex functions, one is instructed to 
remove, by hand, those diagrams in which there is at least one pair of 
external points between which no paths exist consisting solely of 
$\phi$-field propagators.  The corresponding diagrams in the VT theory 
are those in which there is at least one pair of external points between 
which no paths exist consisting solely of propagators having unconstrained 
wave vectors, i.e., freely-unconnected diagrams.  We now show 
that these freely-unconnected diagrams of the VT theory automatically 
vanish in the thermodynamic limit. 

\begin{figure}[hbt]
\epsfxsize=3.5in
\centerline{\epsfbox{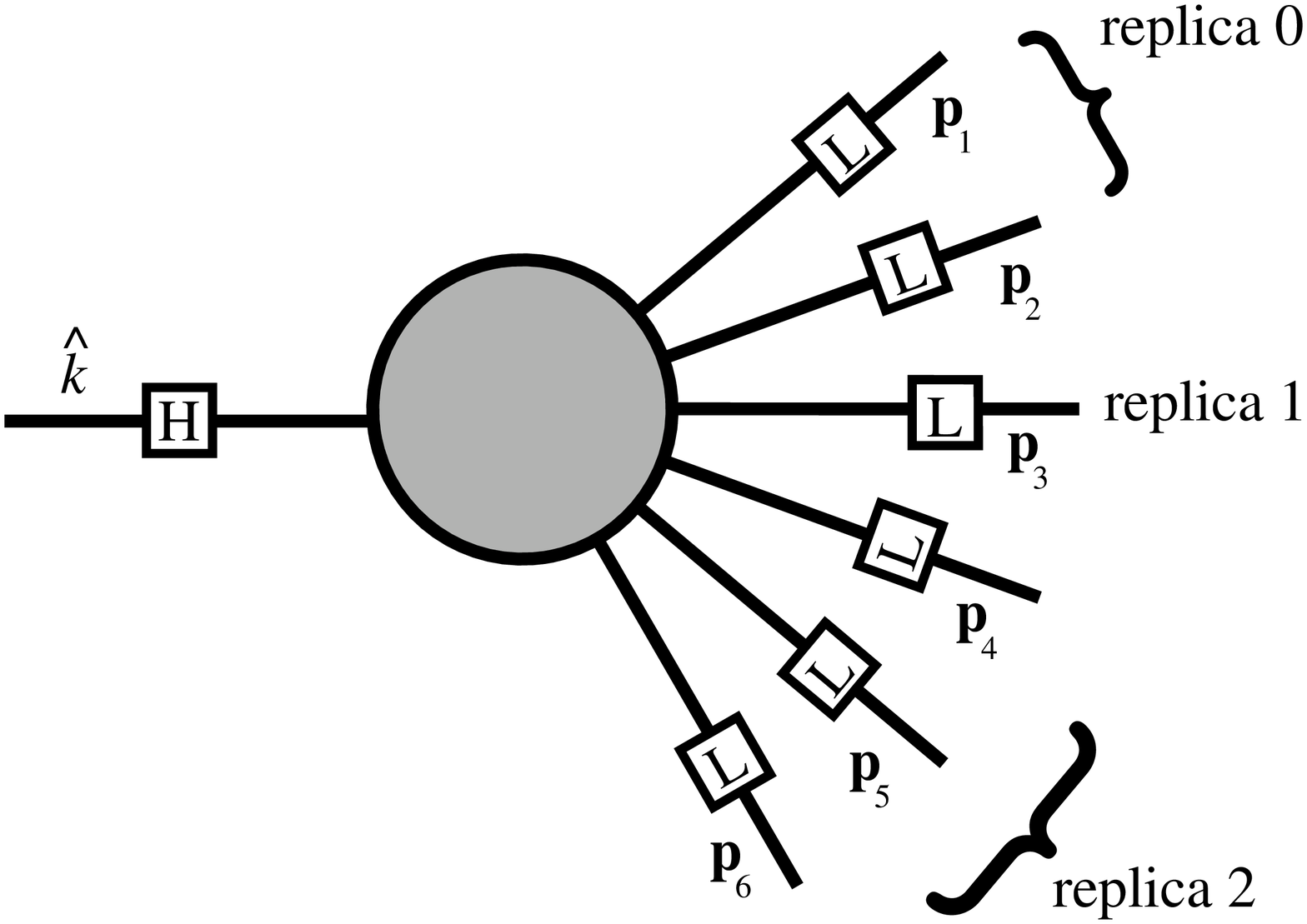}}  
\vskip0.50cm
\caption{Schematic illustration of a piece of a diagram in the 
VT field theory obtained by cutting \LRS\ lines in a freely-unconnected 
diagram.  Note that the wave vector $\hat k$ flowing through the 
external point lies in the \HRS.  The shaded circle represents any way 
to connect the exhibited lines using the cubic interaction vertex, 
unconstrained propagators and \LRS\ propagators.
\label{FIG:Path}}
\end{figure}
To do this, consider a generic VT theory diagram for the two- or 
three-point vertex functions.  Observe that freely-unconnected diagrams 
have the following property: as there exists a pair of external points 
not connected by a path of unconstrained internal lines, there must 
exist at least one scheme of cutting solely \LRS\ internal lines that 
causes the diagram to separate into disconnected pieces with the 
external points shared amongst the pieces.  As we are considering only 
two- and three-point vertex functions, at least one of these pieces
involves only a single \HRS\ external point, along with a number of 
cut \LRS\ lines.  A schematic illustration of such a piece is shown 
in Fig.~\ref{FIG:Path}.

Let us examine the consequences of applying wave-vector conservation 
to this piece, noting that the wave vector ${\hat k}$ flowing in through 
the external point must lie in the \HRS, whereas the wave vectors flowing 
out through the remaining (i.e.~cut) lines lie in the \LRS.  
Now, according to the VT 
field theory, wave-vector conservation requires that the incoming \HRS\ 
wave vector ${\hat k}$ be equal, replica by replica, to the sum of the 
outgoing ($m=2,3,\ldots$) \LRS\ wave vectors flowing in a given 
replica, i.e., that 
\begin{equation}
{\bf k}^{\alpha}=
\sum_{j=1}^{m}\delta^{\alpha,\alpha_{j}}{\bf p}_{j}
\quad
({\rm for}\,\,\alpha=0,1,2,\ldots,n) 
\end{equation}
where $\alpha_{1},\alpha_{2},\ldots$ 
indicate the replicas through which 
wave vectors 
${\bf p}_1,
 {\bf p}_2,\ldots$ flow.  As a consequence, because 
the incoming wave vector lies in the \HRS, the outgoing \LRS\ wave vectors 
must flow out through more than one replica.  This is the key observation, 
as the following example, depicted in Fig.~\ref{FIG:Path} reveals.  Here, 
there are six outgoing \LRS\ lines, 
two   with wave vectors flowing in replica 0, 
one   with wave vector  flowing in replica 1, and 
three with wave vectors flowing in replica 2. 
For replica 0, wave-vector conservation reads
${\bf k}^{0}={\bf p}_{1}+{\bf p}_{2}$, so that, e.g., 
${\bf p}_{1}$ determines ${\bf p}_{2}$. 
Similarly, for replica 1, wave-vector conservation reads 
${\bf k}^{1}={\bf p}_{3}$, so that 
${\bf p}_{3}$ is determined.
More generally, as this special case exemplifies, the number of 
independent outgoing \LRS\ wave vectors is reduced by at least {\it two\/} 
(rather than {\it one\/} that total wave-vector conservation demands) 
simply because of the fact that the outgoing \LRS\ wave vectors 
must flow out through more than one replica.   This, in turn, means 
that in the uncut diagram there are fewer independent wave vectors 
to be summed over than the number of loop wave vectors suggested by 
simple topological counting.  As a result, additional denominators 
of $V^{n+1}$ remain, even after the summation over independent 
wave vectors in the uncut diagram are replaced by their thermodynamic-limit 
integrals, which renders the corresponding freely-unconnected diagram 
negligible.  

As a concrete example of the foregoing argument, we compute the third 
diagram on the right-hand side of the equation depicted in 
Fig.~\ref{FIG:Ex_Decomposition}.  In this diagram, both of the internal 
lines lie in the \LRS, the diagram does not (by simple wave-vector 
conservation) contribute unless the external wave vector $\hat k$ has 
nonzero $d$-vector components in precisely in two replicas 
(e.g.~replicas one and two).  
In this case, the diagram makes the contribution 
\begin{equation}
2V^{n+1}\,G_{0}({\bf k}^1)\,
 V^{n+1}G_{0}({\bf k}^2)\,
 \left({g}\,{V^{-2(n+1)}}\right)^2=
2g^2\,V^{-2(n+1)}\,G_{0}({\bf k}^1)\,G_{0}({\bf k}^2).
\label{EQ:Ex_Cal}
\end{equation} 
On the right-hand side, one denominator of $V^{n+1}$ will combine 
with the Kronecker $\delta$-function to maintain overall 
wave-vector conservation (via a Dirac $\delta$-function in the 
thermodynamic limit); the other  $V^{n+1}$ denominator (which, in 
usual cases, would combine with the summation over a loop wave vector 
to produce an integral) makes this diagram vanish.  This special case 
exemplifies the general emergence, in the VT setting, of the central 
aspect of the HRW formulation, viz., the removal of the 
$\phi$-unconnected diagrams.
\subsection{Replica sums and their decomposition in the replica limit}
\label{SEC:replica}
Now that we have demonstrated that only the freely-connected VT 
field theory diagrams 
contribute, we make a closer examination of these diagrams for the 
relevant cases of the two- and three-point vertex functions.  We begin 
by noting that each diagram exhibits the full symmetry of the VT 
field theory, viz., invariance under separate $d$-dimensional rotations 
in each replica and permutations of the replicas.  Therefore, the
small-wave-vector expansion of $\Gamma_n^{(2)}(\hat k)$ given in 
Eq.~(\ref{EQ:VFTexp}) remains valid, diagram by diagram.

Now, the computation of any contributing diagram involves summations 
over a number of independent \LRS\ (but otherwise unconstrained) wave 
vectors as well as summations over the replicas through which these wave 
vectors flow.  These latter summations over replicas can be 
decomposed as follows:
\begin{equation}
\sum_{\alpha_1,\alpha_2,\ldots,\alpha_l}
F_{\alpha_1,\alpha_2,\ldots,\alpha_l}=
\left(
\sum_{{\alpha_1,\alpha_2,\ldots,\alpha_l
\atop{\rm all\,\, equal}}}+
\sum_{{\alpha_1,\alpha_2,\ldots,\alpha_l
\atop{\rm two\,\, distinct}}}+
\sum_{{\alpha_1,\alpha_2,\ldots,\alpha_l
\atop{\rm three\,\, distinct}}}+\cdots
\sum_{{\alpha_1,\alpha_2,\ldots,\alpha_l
\atop{\rm all\,\, distinct}}}
\right)
F_{\alpha_1,\alpha_2,\ldots,\alpha_l}, 
\label{EQ:Decomposition}
\end{equation}  
where $F_{\alpha_1,\alpha_2,\ldots,\alpha_l}$ is a generic function 
of the $l$ replica indices.  Said equivalently, the summation can be 
decomposed into: terms in which all wave vectors flow through a common 
replica; those in which the wave vectors flow through two distinct 
replicas; etc.

Now let us make use of this decomposition.  
For $\coeffzero$ and $\coeffthree$ in Eq.~(\ref{EQ:VFTexp})
the external wave vector is 
zero~\cite{REF:Wavevectorzero}, and therefore the summand 
$F_{\alpha_1,\alpha_2,\ldots,\alpha_l}$ is invariant under 
permutations of the replicas~\cite{REF:permdef}.
Thus, in the first term of the decomposition, $F$ is constant 
[i.e.~independent of the (common) value of the replica arguments], 
and hence this term contributes $(n+1)F_{0,0,\ldots,0}$. 
In the replica limit, this becomes $F_{0,0,\ldots,0}$.
As for the second term, let us further decompose it into 
partitionings of the set of replica indices into two subsets, 
the replica indices in each subset having a common value.  In 
each such partitioning $F$ is constant, and thus each 
partitioning contributes $(n+1)nF$, which vanishes in the 
replica limit.  By continuing with this decomposition tactic via 
tri-partitioning, tetra-partitioning, etc., we establish that all of 
the terms on the right-hand side of Eq.~(\ref{EQ:Decomposition}) 
except the first vanish in the replica limit.

We next consider the coefficient $\coefftwo$.  
As mentioned above, symmetry considerations dictate that each 
diagram contributing to the two-point vertex function  
has the small-wave-vector expansion 
\begin{equation}
\coeffzero^{\rm (dia)}+
\coefftwo^{\rm (dia)}
{\hat k}\cdot{\hat k}+\cdots\,,
\end{equation}
where $\coeffzero^{\rm (dia)}$ and $\coefftwo^{\rm (dia)}$ are the 
contributions to $\coeffzero$ and $\coefftwo$ from the diagram 
in question.  We exploit the (larger than mandated) $(n+1)d$-dimensional 
rotational invariance of the terms retained in this small-wave-vector
expansion by choosing $\hat k$ to be rotated into a single replica: 
$\{{\bf k},{\bf 0},\ldots,{\bf 0}\}$.  (Although it has, until this 
stage, been vital to ensure that ${\hat k}$ lies in the \HRS, e.g., 
in order to eliminate the freely-unconnected diagrams, one is now at 
liberty to ignore this requirement.)\thinspace\ Repeating the 
tactic just used for the analysis of $\coeffzero$ and $\coeffthree$, 
with the slight elaboration needed to accommodate the fact that 
the (suitably rotated) external wave vector 
$\{{\bf k},{\bf 0},\ldots,{\bf 0}\}$ breaks the permutation 
symmetry group from $\perdef_{n+1}$ down to $\perdef_{n}$.
In this way, we see that the only contributions that survive the 
replica limit are from the {\it all-equal\/} partition and, 
furthermore, from the case in which all of the independent 
\LRS\ wave vectors lie in replica zero.  
\subsection{Feynman integrals and their reduction 
to HRW integrals in the replica limit}
\label{SEC:schwing}
Having shown that the topology and combinatorics of the VT field 
theory diagrams for $\coeffzero$, $\coefftwo$ and $\coeffthree$ 
coincide with those of the HRW field theory diagrams, 
the task that remains is to show that, for 
every diagram contributing to these coefficients, the actual 
{\it value\/} of the corresponding Feynman integral 
reduces, in the replica limit, to the appropriate HRW value.  That 
this is so can most straightforwardly be seen by employing the 
Schwinger representation~\cite{REF:LeBellac} of the 
 powers of the propagator, viz., 
\begin{equation}
\frac{(s-1)!}{(\deltat+\hat{k}\cdot\hat{k})^{s}}=
\int_{\sigma}\sigma^{s-1}\,
{\rm e}^{-\sigma(\deltat+\hat{k}\cdot\hat{k})}=
\int_{\sigma}\sigma^{s-1}
{\rm e}^{-\sigma\deltat}
\prod_{\alpha=0}^{n}
{\rm e}^{-\sigma{\bf k}^{\alpha}\cdot{\bf k}^{\alpha}} 
\qquad\quad({\rm for\/}\,\,\,s=1,2,\ldots),
\label{EQ:jssparam}
\end{equation}
where the Schwinger parameter $\sigma$ ranges between $0$ and 
$\infty$.  Observe that Eq.~(\ref{EQ:jssparam}) presents the propagator 
in a form that is very conveniently factorized on the replica index.  

\begin{figure}[hbt]
\epsfxsize=2.5in
\centerline{\epsfbox{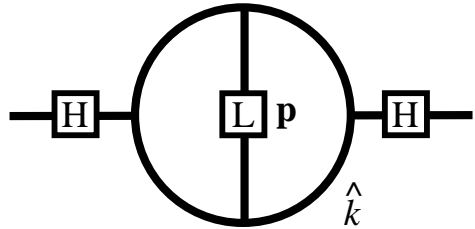}}  
\vskip0.50cm
\caption{A two-loop diagram used to exemplify the decoupling 
of the replicas using the Schwinger representation.
\label{FIG:Schwinger}}
\end{figure}
Let us begin with a concrete example.  In the thermodynamic limit, 
the diagram depicted in Fig.~\ref{FIG:Schwinger} contributes to 
$\coeffzero$ a term proportional to 
\begin{eqnarray}
&&
\inte^{4}\int
d^{(n+1)d}{k}\,\,d^{d}{p}\,\,
G_{0}({\hat k})^{2}\,
G_{0}({\bf p})^{2}\,
G_{0}({\hat k}+{\bf p})^{2}
\nonumber\\
&&\qquad=
\inte^{4}\int
d^{d}{k^{0}}\,\,d^{d}{p}\,\,
\int_{\sigma_{1}\sigma_{2}\sigma_{3}}
\!\!\!\!\!\!\!\!\!
\sigma_{1}\sigma_{2}\sigma_{3}\,
{\rm e}^{-(\sigma_{1}+\sigma_{2}+\sigma_{3})\deltat}
{\rm e}^{-(\sigma_{1}+\sigma_{3}){\bf k}^{0}\cdot{\bf k}^{0}}\,
{\rm e}^{-(\sigma_{2}+\sigma_{3}){\bf p}\cdot{\bf p}}\,
{\rm e}^{-2\sigma_{3}{\bf k}^{0}\cdot{\bf p}}\,
\prod_{\alpha=1}^{n}\int
d^{d}{k^{\alpha}}\,
{\rm e}^{-(\sigma_{1}+\sigma_{3}){\bf k}^{\alpha}\cdot{\bf k}^{\alpha}}
\nonumber\\
&&\qquad=
\inte^{4}\int
d^{d}{k^{0}}\,\,d^{d}{p}\,\,
\int_{\sigma_{1}\sigma_{2}\sigma_{3}}
\!\!\!\!\!\!\!\!\!
\sigma_{1}\sigma_{2}\sigma_{3}\,
{\rm e}^{-(\sigma_{1}+\sigma_{2}+\sigma_{3})\deltat}
{\rm e}^{-(\sigma_{1}+\sigma_{3}){\bf k}^{0}\cdot{\bf k}^{0}}\,
{\rm e}^{-(\sigma_{2}+\sigma_{3}){\bf p}\cdot{\bf p}}\,
{\rm e}^{-2\sigma_{3}{\bf k}^{0}\cdot{\bf p}}\,
\left(\int d^{d}{k}\,
{\rm e}^{-(\sigma_{1}+\sigma_{3}){\bf k}\cdot{\bf k}}
\right)^{n}
\nonumber\\
&&\qquad
\buildrel{n\to 0}\over{\longrightarrow}
\inte^{4}\int
d^{d}{k}\,d^{d}{p}\,
\int_{\sigma_{1}\sigma_{2}\sigma_{3}}
\!\!\!\!\!\!\!\!\!
\sigma_{1}\sigma_{2}\sigma_{3}\,
{\rm e}^{-(\sigma_{1}+\sigma_{2}+\sigma_{3})\deltat}
{\rm e}^{-(\sigma_{1}+\sigma_{3}){\bf k}\cdot{\bf k}}\,
{\rm e}^{-(\sigma_{2}+\sigma_{3}){\bf p}\cdot{\bf p}}\,
{\rm e}^{-2\sigma_{3}{\bf k}\cdot{\bf p}} 
\nonumber\\
&&\qquad
=\inte^{4}\int
d^{d}{k}\,\,d^{d}{p}\,\,
G_{0}({\bf k})^{2}\,
G_{0}({\bf p})^{2}\,
G_{0}({\bf k}+{\bf p})^{2}. 
\end{eqnarray}
This limiting value is {\it precisely\/} that occurring for the 
corresponding diagram in the HRW field theory for the PT. 

The tactic that we have just employed, viz., the use of the Schwinger 
representation to decouple the replicas from one another, provides 
easy and explicit access to the replica limit and, hence, to the 
precise correspondence with the HRW prescription.  It can 
straightforwardly be invoked not only for all diagrams that 
contribute to the coefficients $\coeffzero$ and (by the same 
procedure) $\coeffthree$, but also for the coefficient $\coefftwo$. 

When considering diagrams contributing to $\coeffzero$ and $\coeffthree$, 
we saw that what survived were terms in which all internal \LRS\ 
wave vectors flowed in a common (but otherwise arbitrary) replica. 
Now, as we consider $\coefftwo$, there is a slight complication 
arising from the presence of an external wave vector, which spoils 
the full $\perdef_{n+1}$ permutation symmetry.  However, this 
external wave vector 
has been chosen to lie in replica zero and, as we have shown above, 
the only surviving contribution is the one in which  all internal 
\LRS\ wave vectors also flow in replica zero.  Then, via the Schwinger 
representation of the propagators,  and via factorization on the replica 
indices, we see that the Feynman integral, in the replica limit, is 
identical to that in the HRW approach.  Hence, the VT presents 
not only the the same coefficients $\coeLVzero_{0}$ and 
$\coeLVthree_{0}$ of the two- and three-point vertex functions 
as does the HRW representation, but also the same coefficient 
$\coeLVtwo_{0}$.
\section{Concluding remarks}
\label{SEC:summary}
Let us summarize what is presented in this Paper.  We have addressed 
the vulcanization transition via a minimal field-theoretic model.  
This model is built from an order parameter whose argument is the 
$(n+1)$-fold replication of ordinary $d$-dimensional space.  
[The structure of this theory should be contrasted with that of more 
familiar replica field theories, in which it is the (internal) 
{\it components\/} of the field that are replicated rather than the 
(external) {\it argument\/}]\thinspace\ 
We have considered appropriate long-wave-length 
aspects of the two- and three-point vertex functions for this model, 
to all orders in perturbation theory in the cubic nonlinearity.  
Via a detailed analysis of the diagrammatic expansion for these
quantities, we have found 
that, in the replica limit, these vulcanization-theory vertex functions 
precisely coincide with the corresponding vertex functions of a certain 
field-theoretic representation (due to Houghton, Reeve and Wallace) of 
the percolation transition.  
Hence, percolation theory correctly captures the critical phenomenology 
the liquid and critical states of vulcanized matter.
\section*{Acknowledgments}
\label{SEC:Acknowledgments}
We thank John Cardy, Horacio Castillo and Michael Stone 
for informative discussions.  
This work was supported by the U.S.~National Science 
Foundation through grant DMR99-75187. 
  

\begin{references}
\bibitem{REF:Wavevectorzero}
Strictly speaking, as the wave vector in question is required to 
be in the \HRS, it can not be exactly $\hat 0$.  Instead, one 
should conceive of the value at $\hat 0$ as being the 
limiting value as $\hat k \to \hat 0$ via a path in the \HRS.
\bibitem{REF:permdef}
By this we mean
$F_{\alpha_1,\alpha_2,\ldots,\alpha_l}=
 F_{\pi(\alpha_1),\pi(\alpha_2),\ldots,\pi(\alpha_l)}$, 
where $\pi(\alpha)$ is an arbitrary permutation of the replicas.
\end{references}
\end{document}